\documentclass{article}
\title{
\begin{flushright}
\normalsize hep-ph/0001069 \\
\normalsize TMUP-HEL-9907\\
\end{flushright}
Matter effects on neutrino oscillations in
gravitational and magnetic fields}
\author{H. Athar\\
Department of Physics, Tokyo Metropolitan University,\\
Minami-Osawa 1-1, Hachioji, Tokyo 192-0397,\\
Japan\\[12pt]
Jos\'e F. Nieves\\
Laboratory of Theoretical Physics \\
Department of Physics, P.O. Box 23343 \\
University of Puerto Rico, R\'{\i}o Piedras \\
Puerto Rico 00931-3343
}
\date{June 1999}

\begin{document}
\maketitle

\begin{abstract}

When neutrinos propagate in a background, 
their gravitational couplings are modified 
by their weak interactions with the
particles in the background.
In a medium that contains electrons but no muons or taons, 
the matter-induced gravitational couplings of neutrinos
are different for the various neutrino flavors, and
they must be taken into account in describing
the phenomena associated with the neutrino oscillations in the presence
of strong gravitational fields.  Here we incorporate
those couplings in that description, 
including also the effects of a magnetic field,
and consider the implications that they have for
the emission of high energy neutrinos in the vicinity of Active
Galactic Nuclei.

\end{abstract}

%
% section 1
%
\section{Introduction}
\setcounter{equation}{0}

It is now well known that the interactions of a neutrino
with the background particles
can influence the neutrino properties in essential ways as it
propagates through a medium.  Those effects can have
important consequences for various physical phenomena,
such as the Mikheyev-Smirnov-Wolfenstein (MSW) mechanism and its
variations in the context of the solar neutrino problem,
the explanation of the large velocities of pulsars
in terms of the asymmetric emission of neutrinos from the
proto neutron star and the propagation of high energy 
($E\geq 10^{6}$ GeV) neutrinos in the vicinity of cores of 
Active Galactic Nuclei (AGN) \cite{R0}.

In all these applications, a common ingredient is the fact that
the various neutrino flavors interact with the background
particles differently.   
While the standard neutrino interactions in the fundamental
Lagrangian have a universal character, the universality
is broken when the effects of the medium are taken into account.

In the presence of a medium, these
breakdown of the universality of the neutrino interactions
includes also the gravitational ones.  
In Ref.\ \cite{np:gravnu} the effective gravitational vertex
of neutrinos was determined by calculating the one-loop
contribution to the neutrino stress-energy tensor.
In the presence of a static gravitational potential,
such matter-induced interactions lead to an additional
contribution to the neutrino dispersion relations,
or equivalently their indices of refraction, which
depend on the gravitational potential and
are not the same for all the neutrino flavors.
As was stressed in that work, those calculations were
based on the standard model of particle interactions
and the commonly accepted linearized theory of gravity
with a universal tree-level coupling.  The non-universal
character of the induced gravitational couplings is a
consequence of the flavor asymmetry of the background
and not of an assumed breakdown of universality
of the fundamental Lagrangian.

On the other hand, it has been observed by P\'{\i}riz, Roy and 
Wudka \cite{prw},
that high energy neutrinos originating from AGN can
have resonant spin-flavor transitions due to the combined
effects of the gravitational interactions and the presence
of a large magnetic field.  AGN can be a source of
high energy neutrinos and, therefore, a good understanding
of such transitions is useful to estimate the expected fluxes
from these objects in the forthcoming searches for high
energy neutrinos in neutrino telescopes.

In the calculations of Ref.\ \cite{prw}, the matter, magnetic and gravitational
effects were considered.  However, the latter were introduced
using the standard tree-level couplings of neutrinos to gravity and,
in particular, the effects of the medium on the 
effective gravitational interactions were not taken into account.
This amounts to neglect the additional
matter-induced gravitational contribution to the neutrino
dispersion relation. As we have already mentioned, they
depend on the gravitational potential and, 
what is more important for the issue of resonant transitions,
are not flavor symmetric.  In contrast, the tree-level gravitational
couplings
are the same for all the neutrino flavors, including the
right handed (singlet) neutrinos, and hence should have no
effect whatsoever on the phenomenon of neutrino oscillations.

Motivated by these considerations, in this work we
take another look at the subject of neutrino oscillations
in the presence of a gravitational field. Our objective
is to pay particular attention to the effects of the
matter-induced gravitational couplings that we have
already mentioned, with a view to their implications
in the context of high energy neutrino emission from AGN.

We start in Section \ref{sec:gravnu} with a brief
overview of the main results obtained
in Ref.\ \cite{np:gravnu}  that are
needed in subsequent sections here.  
This includes a summary of the formulas obtained
there for the dispersion relations for standard left-handed
neutrinos in the presence of a gravitational potential,
taking into account the matter-induced terms in the gravitational
vertex of the neutrino.
In Section \ref{sec:general} we set up in general terms
the equations that are relevant for treating 
the phenomenon of resonant neutrino (spin)-flavor transitions
in the presence of a gravitational potential. 
The treatment includes the possibility that a 
magnetic field may also be present, 
under the assumption that the neutrinos have an intrinsic magnetic
moment.  Using the results of Section \ref{sec:general}
as the starting point, in Section \ref{sec:agn} we study the effect
of the resonant transitions on the
the determination of the flux of high energy neutrinos
emitted by AGN, and finally the conclusions are given in 
Section \ref{sec:conclusions}.

%
% section 2
%
\section{Matter effects on the neutrino gravitational interactions}
\setcounter{equation}{0}
\label{sec:gravnu}

In Ref.\ \cite{np:gravnu} we obtained the dispersion relation
that is obeyed by a standard left-handed neutrino in the presence
of a static gravitational potential. In this section
we review briefly those results and consider the corresponding ones
for a right-handed (singlet) neutrino .

As already indicated, those results
were derived in the linearized theory of gravity, in which the metric
tensor is written as
\begin{eqnarray}
g_{\lambda\rho} = \eta_{\lambda\rho} + 2\kappa h_{\lambda\rho},
\label{defh}
\end{eqnarray}
where $\eta_{\lambda\rho}$ is the flat space metric. We then expand the
Lagrangian in the presence of gravity in powers of $\kappa$ and keep only
the first order terms. In this formulation, $h_{\lambda\rho}$ appears as
the graviton field, which is a spin-2 quantum field coupled to the
stress-energy tensor, whose interactions can be studied in the flat
Minkowskian background. The Einstein-Hilbert action for pure gravity is
given by
\begin{eqnarray}
{\cal A} = {1\over 16\pi G} \int d^4x\; \sqrt{-{\tt g}} \; R \,,
\end{eqnarray}
where $R$ is the Ricci scalar, $\tt g$ is the determinant of the
matrix $g_{\lambda\rho}$, and $G$ is the Newton's constant. Using Eq.\
(\ref{defh}), we can verify that this gives the correct kinetic terms
for the spin-2 field if we make the identification
\begin{eqnarray}
\kappa = \sqrt{8\pi G} \,.
\end{eqnarray}
Then starting from the Dirac Lagrangian for a given fermion $f$
in the presence of gravity, the 
coupling of the graviton field $h_{\lambda\rho}$ with the fermion 
field can be written as
\begin{eqnarray}\label{Leeh}
{\cal L}^{(ff)}_{h} = -\kappa h^{\lambda\rho} (x) \widehat
T^{(f)}_{\lambda\rho} (x) \,, 
\end{eqnarray}
where the stress-energy tensor operator $\widehat
T_{\lambda\rho}^{(f)}$ for the fermion field is given by
\begin{eqnarray}\label{stresstensor}
\widehat T^{(f)}_{\lambda\rho} (x) = \left\{
{i\over 4} \overline \psi(x) \left[\gamma_\lambda \partial_\rho + \gamma_\rho
\partial_\lambda \right] \psi(x) + H.c. \right\} -
\eta_{\lambda\rho} {\cal L}_0^{(f)} (x) \,.
\end{eqnarray}
Here ${\cal L}_0^{(f)}(x)$ is the Lagrangian for the
free Dirac field, which we write in the explicitly Hermitian form
\begin{eqnarray}\label{Lf0}
{\cal L}^{(f)}_0 = \left[
\frac{i}{2}\overline\psi \gamma^\mu \partial_\mu \psi 
+ H.c. \right]
- m_f \overline\psi\psi  \,.
\end{eqnarray}
{}From Eqs.\ (\ref{Leeh}) and (\ref{stresstensor}) 
it follows that the term associated
with the gravitational fermion vertex in a Feynman diagram
is $-i\kappa V^{(f)}_{\lambda\rho}$, where
\begin{eqnarray}\label{V}
V^{(f)}_{\lambda\rho} (p,p') = \frac14 \left[
\gamma_\lambda(p + p')_\rho + 
\gamma_\rho(p + p')_\lambda \right]
- \frac12 \eta_{\lambda\rho}
\left[(\rlap/ p - m_f) + (\rlap/ p' - m_f) \right] \,.
\end{eqnarray}
It is easy to deduce by inspection that for a left ($\nu_L$) 
or a right ($\nu_R$) handed massless neutrino, 
the corresponding quantity is given by
\begin{eqnarray}
\label{VnuLR}
V^{(\nu_{L,R})}_{\lambda\rho} (k,k') = \frac14 \left[
\gamma_\lambda(k + k')_\rho + 
\gamma_\rho(k + k')_\lambda \right]\chi
- \frac12 \eta_{\lambda\rho}
\left[\rlap/ k  + \rlap/ k' \right]\chi \,,
\end{eqnarray}
where $\chi = L,R \equiv \frac{1}{2}(1 \mp \gamma_5)$.

When the effects of the background medium are taken
into account, the result is that the matrix element
of the total stress-energy tensor $\widehat T_{\mu\nu}(x)$
between neutrino states,
with incoming and outgoing momenta $k$ and $k^\prime$
respectively, is given by
\begin{equation}\label{defGamma}
\langle\nu_{L,R}(k^\prime)|\widehat T_{\mu\nu}(0)|\nu_{L,R}(k)\rangle
= \overline u_{L,R}(k^\prime)\left(V^{(\nu_{L,R})}_{\mu\nu}(k,k^\prime)
+ \Lambda^{(\nu_{L,R})}_{\mu\nu}(k,k^\prime)\right)
u_{L,R}(k)\,.
\end{equation}
The quantity $\Lambda^{(\nu_{L,R})}_{\mu\nu}(k,k^\prime)$ represents
the matter-induced contribution and its calculation
to one-loop was the subject of Ref.\ \cite{np:gravnu}.
That calculation was performed for the left-handed massless
neutrinos, adopting the standard electro-weak couplings
for the neutrinos and the other particles. The corresponding result
for the right-handed (singlet) neutrinos, since they have no standard
couplings to matter, is
\begin{equation}\label{LamdaR}
\Lambda^{(\nu_R)}_{\mu\nu} = 0 \,.
\end{equation}

The particular formulas for $\Lambda^{(\nu_L)}_{\mu\nu}$ are
not relevant for us here. More important is the result
that they imply for the dispersion relation that a
neutrino obeys in the presence of a static gravitational
potential $\phi^{ext}$.  Using Eq.\ (\ref{defGamma}) as the starting point,
it was shown in Ref.\ \cite{np:gravnu} that the dispersion
relation of a standard left-handed neutrino that propagates 
with momentum $\vec K$ is given by
\begin{equation}\label{omegaL}
\omega^{(L)}_K = K + 2K\phi^{ext} + b_{\mbox{mat}} + b_G \,,
\end{equation}
while for its antiparticle (a right-handed antineutrino) the sign
in front of $b_{\mbox{mat}}$ and $b_G$ is the opposite.
The coefficient $b_{\mbox{mat}}$ is the usual Wolfenstein term while $b_G$
represents the matter-induced gravitational contribution.  
Following the same arguments, it follows that the dispersion
relation $\omega^{(s)}_K$
for a sterile neutrino (either left- or right-handed) is simply
\begin{equation}\label{omegaR}
\omega^{(s)}_K = K + 2K\phi^{ext} \,.
\end{equation}

It is therefore clear that the tree-level gravitational neutrino
couplings, which are represented in Eqs.\ (\ref{omegaL}) and (\ref{omegaR}) by
the $2K\phi^{ext}$ term, do not play any role whatsoever in
the phenomenon of neutrino oscillations, since they 
appear as a common factor in the dispersion relations of
all the neutrino states.
In contrast, the matter-induced terms, which are zero
for the right-handed singlet neutrinos and do not have the same value
for the three left-handed neutrino flavors, affect the
neutrino oscillations via the MSW mechanism\cite{footnote1}.  In those
situations in which the gravitational contribution can be
neglected, Eq.\ (\ref{omegaL}) reduces simply to the Wolfenstein
formula. However, in environments in which a relatively strong gravitational
potential is present, the matter-induced  gravitational term
is important and
must be taken into account in any discussion of MSW-type
effects under such conditions.  This is the subject 
we take up next.

%
% section 3
%
\section{Oscillations in a gravitational and a magnetic field}
\setcounter{equation}{0}
\label{sec:general}

We consider two neutrino families, the first of which we take
to be $\nu_e$, and denote the second one by $\nu_x$, which
can be either one of $\nu_{\mu,\tau}$ or a sterile neutrino $\nu_s$.
Each family consists of a 
left-handed neutrino field $\nu_{a L} (a = e,x)$, and a corresponding 
right-handed partner $N_{a R}$ that we take to be a weak singlet.
In addition to the mixing via the mass matrix, we allow the
possibility that the neutrinos have an intrinsic magnetic moment 
coupling.  In situations in which a magnetic field is present,
those couplings can induce the (resonant) spin transitions 
between the two neutrino helicity states.
We consider two situations separately, according to the
type of the magnetic moment coupling that they may have.

\paragraph{Case I}

Here we assume that the neutrinos have a magnetic
moment interaction defined by the term
\begin{equation}\label{defmu}
L^\prime = -\frac{1}{2}\sum_{a,b}\mu_{ab}\bar N_{a R}
\sigma_{\mu\nu}\nu_{b L}F^{\mu\nu} + H.c. \,,
\end{equation}
in the Lagrangian. As shown in the Appendix, if the neutrino flavor
amplitudes are assembled in the vector 
\begin{equation}\label{Chi}
\chi = 
\left(
\begin{array}{c}
\alpha_{\nu_{eL}} \\
\alpha_{\nu_{xL}} \\
\beta_{\nu_{eR}} \\
\beta_{\nu_{xR}}
\end{array}
\right),
\end{equation}
then their evolution 
is governed by the Hamiltonian matrix
\begin{equation}\label{HamB}
H = 
\left(\begin{array}{cc}
K + b_{\mbox{mat}} + b_G + 
\frac{(m + \mu B_{||})^2}{2K} & -\mu B_T \\[12pt]
-\mu B_T & K + \frac{(m - \mu B_{||})^2}{2K}
\end{array}
\right),
\end{equation}
where each entry in this equation is itself a $2\times 2$ matrix.
In particular, $m$ is the neutrino mass mixing matrix and
$\mu$ is the magnetic moment matrix defined in Eq.\ (\ref{defmu}),
while $K$ denotes the magnitude of the neutrino momentum
and $B_{||}, B_T$ are the components of the magnetic
field parallel and transverse to $\hat K$, respectively.
The terms $b_{\mbox{mat}}$ and $b_G$ are diagonal matrices
whose values depend on the composition of the background
medium.  At this point it is useful to introduce
the particle and antiparticle momentum distribution functions
\begin{eqnarray}
\label{fe}
f_{f,\bar f}(p_f) = \frac{1}{e^{\beta(E_f \mp \mu_f)} + 1}\,, 
\end{eqnarray}
where the upper and the lower signs hold for the particle and the
antiparticle respectively, 
$\mu_f$ is the chemical potential of the
fermion $f$ and
\begin{eqnarray}\label{pErel}
p^\mu_f = (E_f,\vec P)\,, \qquad E_f = \sqrt{\vec P^2 + m_f^2} \,.
\end{eqnarray}

The corresponding total number densities
are given by
\begin{equation}\label{ne}
n_f = 2 \int {d^3P \over (2\pi)^3} \; f_f \,, \qquad 
n_{\bar f} = 2 \int {d^3P \over (2\pi)^3} \; f_{\bar f} \,,
\end{equation}
In terms of these quantities, 
\begin{eqnarray}
\label{bnuall}
b_{\mbox{mat}} = \left\{
\begin{array}{lll}
b_e + \sum_{f} X_f b_f  & \mbox{for $\nu_e$} \\[12pt]
\sum_{f} X_f b_f & \mbox{for $\nu_\mu,\nu_\tau$} \\[12pt]
0 & \mbox{for $\nu_s$},
\end{array}\right. 
\end{eqnarray}
and
\begin{eqnarray}\label{bGfinal}
b_G & = & \phi^{\rm ext} \sqrt{2}G_F \times
\left\{ \begin{array}{lll}
J_e +
\sum_f X_f J_f  & \mbox{for $\nu_e$} \\[12pt]
\sum_f X_f J_f   & \mbox{for $\nu_\mu,\nu_\tau$} \\[12pt]
0 &  \mbox{for $\nu_s$},
\end{array}\right.
\end{eqnarray}
where
\begin{eqnarray}\label{bfJf}
b_f & = & \sqrt{2}G_F(n_f - n_{\overline f}), \nonumber\\
J_f & = & -3(n_f - n_{\overline f}) +
\int\frac{d^3P}{(2\pi)^3 2E_f} \; \frac{dF_f}{dE_f} \,,
\end{eqnarray}
with
\begin{equation}\label{F_f}
F_f = 4(2E_f^2 - m_f^2)(f_f - f_{\overline f}) \,.
\end{equation}
In Eqs.\ (\ref{bnuall}) and (\ref{bGfinal}), the sum over $f$ must be made with respect
to all the particle species that compose the background, and
$X_f$ stands for their neutral current couplings. So, for example,
for a background composed of electrons, neutrons, protons 
and their antiparticles ($f = e,n,p$),
\begin{eqnarray}\label{Zcouplings}
-X_e = X_p  & = & \frac{1}{2} - 2\sin^2\theta_W, \nonumber\\
X_n & = & -\frac{1}{2} \,.
\end{eqnarray}
The explicit formulas for $J_f$ were given in Ref.\ \cite{np:gravnu}
for various limiting cases.

\paragraph{Case II}

In this case we assume that the magnetic moment interaction term
is of the form\cite{footnote:nuc}
\begin{equation}\label{etaumixing}
L^\prime = -\frac{1}{2}\mu\bar\nu^c_{x R}
\sigma_{\mu\nu}\nu_{eL}F^{\mu\nu} + H.c. \,,
\end{equation}
but that the mass terms in the Lagrangian are diagonal, i.e.,
\begin{equation}\label{model2}
L_m = -m_{\nu_e}\bar N_{eR}\nu_{eL} - 
m_{\nu_x}\bar N_{xR}\nu_{xL}
+ H.c. \,,
\end{equation}
so that there is no mixing in the mass matrix.
Denoting by $\alpha_{\nu_e}$ and $\beta_{\bar\nu_x}$ the
$\nu_{eL}$ and $\nu^c_{xR}$ components of the
wave function, and writing them in the form
\begin{equation}\label{Chietau}
\chi = 
\left(
\begin{array}{c}
\alpha_{\nu_{eL}} \\
\beta_{\bar\nu_{xR}}
\end{array}
\right) \,,
\end{equation}
the Hamiltonian matrix that determines their evolution
is given in this case by\cite{footnote:majorana}
\begin{equation}\label{Hametau}
H = 
\left(\begin{array}{cc}
K + b^{(\nu_e)}_{\mbox{mat}} + b^{(\nu_e)}_G + 
\frac{m_{\nu_e}^2}{2K} + \frac{{\mu^2 B_{||}}^2}{2K} & -\mu B_T \\[12pt]
-\mu B_T & K - b^{(\nu_x)}_{\mbox{mat}} - b^{(\nu_x)}_G + 
\frac{m_{\nu_x}^2}{2K} + \frac{{\mu^2 B_{||}}^2}{2K}
\end{array}
\right) \,,
\end{equation}
where $b^{(\nu_a)}_{\mbox{mat}}$ and $b^{(\nu_a)}_G$
$(a = e,x)$ are given in Eqs.\ (\ref{bnuall}) and (\ref{bGfinal}).

Notice that, in the MSW mechanism involving only standard
left-handed neutrinos, the matter effects due to the 
neutral current interactions are the same for all the neutrino species
and therefore are not relevant as far as the oscillation mechanism
is concerned. In contrast, in the situation we have just considered,
those neutral current contributions have opposite sign for $\nu_{eL}$
and $\bar\nu^c_{\mu,\tau R}$, and are zero for $\nu_s$, and 
therefore must be taken into account.
In the physical settings that we are considering,
they can be important since they involve the nucleons and their
couplings to gravity.

Eqs.\ (\ref{HamB}) and (\ref{Hametau}) 
set the framework for our consideration of the
possible effects that the matter-induced gravitational
interactions of the neutrinos may have
for the phenomenon of (resonant) transitions in the
presence of a strong gravitational field.
As a concrete example, we consider those transitions
in the environment of AGN.

%
% section 4
%
\section{Resonant transitions in the vicinity of AGN}
\setcounter{equation}{0}
\label{sec:agn}

For concreteness, we focus our attention on possible transitions between
the $\nu_{e}$ and $\nu_{\tau}$ flavors for the following reason.
Some galaxies have relatively bright centers (as compared to the total 
photon luminosity of the whole galaxy). High energy photons 
reaching tens of thousands of GeV have been observed from these centers. 
If Fermi mechanisms are responsible for accelerating
the electrons in these systems, then
protons are also expected to be accelerated by similar ones. 
If this is true, neutrinos are expected to be produced in $p \gamma$ or/and 
$pp$ collisions, providing the signature for proton acceleration. 
Tau neutrinos are also to be produced in the same collisions but at 
highly suppressed levels, because both the branching ratios and the 
production rates for unstable hadrons that decay into tau neutrinos are much 
lower than for muon or electron neutrinos. These two (suppression) factors 
enter into the calculation of relevant intrinsic neutrino fluxes 
multiplicatively, implying that the intrinsic $\nu_{\tau}$ 
 (and $\bar{\nu}_{\tau}$) flux is relatively rather small. This $\nu_{\tau}$
(and $\bar{\nu}_{\tau}$) flux may possibly be somewhat enhanced due to 
(resonant) transitions between relevant neutrino states. 
     
Briefly, in $p\gamma$ collisions, the  protons and photons may give rise
to high energy $\nu_{\tau}$ (and $\bar{\nu}_{\tau}$)  
mainly through $p+\gamma \, \rightarrow \, D^{+}_{S}+\Lambda^{0}+\bar{D}^{0}$
in addition to producing $\nu_{e}$ and $\nu_{\mu}$ mainly through 
$p+\gamma \rightarrow \Delta^{+}\rightarrow n+\pi^{+}$. 
The production cross-section for $D^{+}_{S}$ is essentially up to three
orders of magnitude lower than that of $\Delta^{+}$ production 
for the relevant center of mass energy scale. 
Moreover the branching ratio of $D^{\pm}_{S}$ to decay eventually into 
$\nu_{\tau}$ ($\bar{\nu}_{\tau}$) is approximately two orders of magnitude 
lower than for $\Delta^{+}$ to subsequently decay into $\nu_{e}$ and 
$\nu_{\mu}$ through $\pi^{+}$. These 
two suppression factors along with the relevant kinematic limits give 
approximately the ratio of intrinsic fluxes of tau neutrinos and 
electron neutrinos as 
$F^{0}(\nu_{\tau}+\bar{\nu}_{\tau})/F^{0}(\nu_{e}+\bar{\nu}_{e})\,
\sim \, 10^{-5}$.

In $pp$ collisions, the $\nu_{\tau}$ flux may be obtained through 
$p+p\rightarrow D^{+}_{S}+X$.
The relatively small cross-section for $D^{+}_{S}$ production 
together with the low branching 
ratio into $\nu_{\tau}$ implies that the 
$\nu_{\tau}$ flux in $pp$ collisions is also suppressed up to $4-5$ 
orders of magnitude relative to $\nu_{e}$ and/or $\nu_{\mu}$ fluxes. 

Thus, in both type of collisions, the estimated intrinsic
$\nu_{\tau}$ flux from the AGNs
is rather small relative to $\nu_{e}$ and/or $\nu_{\mu}$ fluxes,  
essentially for the entire relevant 
neutrino energy range ($2\cdot 10^{6}\leq K$/GeV $\leq 2\cdot 10^{7}$)
\cite{ARZ}. 
However, as we point out
later in this Section, due to transitions between relevant 
neutrino flavors, it is possible that   
$F^{0}(\nu_{\tau}+\bar{\nu}_{\tau})/F^{0}(\nu_{e}+\bar{\nu}_{e})\gg 10^{-5}$,
thus raising the possibility of detection of high energy $\nu_{\tau}$ in
new km$^{2}$ surface area neutrino telescopes within the relevant 
neutrino energy range \cite{LP}.

We first briefly describe the matter density and magnetic field 
profiles in the AGN.
According to \cite{SP}, the matter density profile in the AGN is given
as a function of the distance $r$ from the center 
by
\begin{equation}\label{rho}
\rho_{AGN}(r) = \rho_{0}f(x),
\end{equation}
where $\rho_{0}\sim 1.4\cdot 10^{-12}$ g cm$^{-3}$
(we take the typical luminosity of
the AGN to be $\sim 10^{45}$ erg s$^{-1}$) and $f(x)= x^{-2.5}(1-0.1x^{0.31})
^{-1}$ for $x \sim (10-100)$. Here $x\equiv r/R_{S}$, where $R_{S}$ is the
Schwarzchild radius $R_{S}\simeq 3\cdot 10^{11} 
\left(\frac{M_{AGN}}{10^{8}
M_{\odot}}\right)$m, and in terms of it the gravitational potential
is expressed as $\phi^{ext} = -1/2x$.
We consider the following magnetic field profile in the 
AGN \cite{SP}
\begin{equation}\label{magnet}
 B_{AGN}=B_{0}g(x),
\end{equation}
where $B_{0}\sim 1.7\cdot 10^{5}$ G and $g(x)=x^{-1.75}(1-0.1x^{0.31})^{-0.5}$.
We will use these matter density and magnetic field profiles in our estimates
as an example.

In what follows, the discussion of resonant transitions
in the AGN is divided in two parts, according to whether or 
not the magnetic field terms are important.
We discuss the two necessary conditions for resonant 
transitions to occur in these two situations,
namely, the
level crossing and the adiabaticity at the level crossing, 
and we also comment briefly on the possibility of 
transitions between active and sterile neutrinos.
    
%
%Subsection 4.1
%

\subsection{No magnetic field interactions}
\label{subsec:NoB}

In this situation, the mixing occurs via the mass matrix only. 
After subtracting a term proportional to the identity matrix
in Eq.\ (\ref{HamB}), the relevant 2$\times $2 Hamiltonian matrix 
in the $(\nu_{eL},\nu_{\tau L})$ basis is given by
\begin{equation}\label{HF}
 H_F = \frac{1}{2}
 \left(\begin{array}{cc}
 \Delta H_F & (\Delta m^{2}_{e\tau}/2K)\sin 2\theta_{e\tau} \\[12pt]
 (\Delta m^{2}_{e\tau}/2K)\sin 2\theta_{e\tau} & - \Delta H_F
 \end{array}\right) \,,
\end{equation}
with
\begin{equation}\label{flv}
 \Delta H_{F} = \Delta V_{F}-\frac{\Delta m^{2}_{e\tau}}{2K}\cos 
 2\theta_{e\tau},
\end{equation}
where $\theta_{e\tau}$ is the vacuum mixing angle, 
$\Delta m^{2}_{e\tau} = m^{2}_{\nu_{2}}-m^{2}_{\nu_{1}}$
is the mass squared difference between the two neutrino mass
eigenstates $\nu_{iL} (i = 1,2)$,
and
\begin{equation}\label{VF}
\Delta V_{F}\simeq \sqrt{2}G_{F}(n_{e} + J_{e}\phi^{ext})\,.
\end{equation}
In contrast to the term proportional to the density $n_e$ in
Eq.\ (\ref{VF}), the term proportional to the gravitational potential
depends on the nature of the electron gas.
For the case of a  classical
non-relativistic ($\beta^{-1}\equiv T\ll m_{e})$ electron 
background \cite{np:gravnu},
\begin{equation}\label{Je}
 J_{e}=-\beta m_{e}n_{e}.
\end{equation}
In the following estimates we use
$T\simeq (1-10)$ eV as an example, which is the value that corresponds to 
the black body (thermal) spectrum temperature for the photons
mentioned above \cite{SP}.
This value of $T$ may be considered as the average temperature for the
remaining background particles as well because of thermal equilibrium
(it is mainly this ultraviolet bump in photon 
spectrum that interacts with the 
Fermi accelerated protons producing the relevant unstable hadrons).    

Using the matter density profile given by Eq.\ (\ref{rho}) and for 
$K\sim 7\cdot 10^{6} $ GeV, the level crossing condition 
($\Delta H_{F}=0$) can
be satisfied for $\Delta m^{2}_{e\tau }\sim (10^{-10}-10^{-11})$ eV$^{2}$. 
Without taking into account the (matter induced)
gravity effects, the level crossing condition can also be
satisfied but for values of $\Delta m^{2}_{e\tau}$ that are approximately  
two orders of magnitude smaller than the one just quoted.
However, the other essential condition, namely, the adiabaticity at the level 
crossing\cite{MS}
\begin{equation}\label{KF}
 \kappa_{F}\equiv \left(\frac{\Delta m^{2}_{e\tau}}{2K}\right)^{2}
 \frac{2}{\pi}
 \frac{\sin^{2}2\theta_{e\tau}}{|\Delta \dot{V}_{F}|}\geq 1,
\end{equation}
(where $\Delta \dot{V}_{F}\equiv \mbox{d}(\Delta V_{F})/\mbox{d}r$)
is not satisfied; that is, $\kappa_{F}\ll 1$.
Therefore, the matter induced gravity effects do not
lead to resonant flavor transitions between 
$\nu_{e}$ and $\nu_{\tau}$\footnote{For the channel 
$\nu_{\mu}\rightarrow \nu_{\tau}$ the 
matter effects are absent, but 
there is an interesting possibility of vacuum flavor
oscillations as supported by the recent superkamiokande data \cite{SKK}. 
The vacuum flavor oscillation probability expression
is given by the familiar expression
\[
 P(\nu_{\mu}\rightarrow \nu_{\tau})=\sin^{2}2\theta_{\mu \tau}
 \sin^{2}\left(\frac{\Delta m^{2}_{\mu \tau}}{4K}L\right).
\]
where $\theta_{\mu \tau}$ is the relevant vacuum mixing angle with 
the corresponding mass squared difference
$\Delta m^{2}_{\mu \tau}$. We take $\Delta m^{2}_{\mu \tau}$
and $\sin^{2}2\theta_{\mu \tau}$ values ($\Delta m^{2}_{\mu \tau}\sim 10^{-3}
$ eV$^{2}$ and $\sin^{2}2\theta_{\mu \tau}\sim 1$) as suggested by recent 
superkamiokande data concerning the deficit of atmospheric muon 
neutrino \cite{SKK}. 
Taking the typical distance between the AGN and our galaxy as $L\sim 
100$ Mpc (where 1 pc $\sim 3\cdot 10^{16}$ m), it follows that
$P(\nu_{\mu}
\rightarrow \nu_{\tau}$)$\sim $ 1/2 for $2\cdot10^{6}\leq K$/GeV $\leq 2\cdot
10^{7}$, thus yielding $F(\nu_{\tau}+\bar{\nu}_{\tau})\sim 
 F(\nu_{\mu}+\bar{\nu}_{\mu})$. 
Similarly, vacuum flavor oscillations between $\nu_{e}$ and 
$\nu_{\tau}$ may also take place for certain range of neutrino mixing
parameters resulting in $P(\nu_{e}\rightarrow \nu_{\tau})\sim \sin^{2}2
\theta_{e\tau }$.}.

%
%Subsection 4.2
%

\subsection{With magnetic field interactions}
\label{subsec:withB}

There are several possible cases that can be considered, depending
on the mass mixing matrix and the magnetic moment couplings.
We consider the specific situation in which the 
$\nu_{eL}$ and $\nu^c_{\tau R}$ fields
are mixed by a transition magnetic moment interaction without
mixing in the mass matrix, as described in Case II
in Section \ref{sec:general}.
In this case, after subtracting the term proportional to
the identity in Eq.\ (\ref{Hametau}) 
the relevant Hamiltonian matrix 
in the $(\nu_{eL},\nu^c_{\tau R})$ basis can be written as
\begin{equation}\label{HSF}
 H_{SF} = 
 \frac{1}{2}\left(\begin{array}{cc}
 \Delta H_{SF} & -2\mu B_T \\[12pt]
 -2\mu B_T & - \Delta H_{SF}
 \end{array}\right),
\end{equation}
where
\begin{equation}\label{spin}
\Delta H_{SF} = \Delta V_{SF} - \frac{\Delta m_{e\tau}^2}{2K},
\end{equation}
with
\begin{equation}\label{VFS}
 \Delta V_{SF} \simeq  \sqrt{2}G_{F}\left[
 (1 + 2X_e)(n_{e} + J_{e}\phi^{ext}) + 
 2\sum_{f = n,p} X_f\left(n_f + J_f\phi^{ext}\right)
 \right]
\end{equation}
and $\Delta m_{e\tau}^2 = m_{\nu_\tau}^2 - m_{\nu e}^2$.
For simplicity of the notation, in the reminder of this section 
we take a purely transverse magnetic field, so that $B_T = B$. 
In the more general case, the formulas given below hold with the 
replacement $B \rightarrow B_T$.
Using then Eq.\ (\ref{HSF}), the spin-flavor precession 
probability for constant $B$ and $\Delta V_{SF}$ is given by
\begin{equation}\label{PSPIN}
 P(\nu_{e}\rightarrow \bar{\nu}_{\tau})=
 \left[\frac{(2\mu B)^{2}}{(2\mu B)^{2}+(\Delta H_{SF})^{2}}\right]
 \sin^{2}\left[\sqrt{(2\mu B)^{2}+(\Delta H_{SF})^{2}}\frac{\Delta r}{2}
 \right].
\end{equation}
Here $\Delta r$ is the width of the region where $B$ is appreciable.
Let us study the various interesting situations that may arise from this 
expression for $P(\nu_{e}\rightarrow \bar{\nu}_{\tau})$ under different 
physical conditions.

Let us consider first the case in which the contribution of the
$\Delta V_{SF}$ term in Eq.\ (\ref{PSPIN}) is small. If
$\Delta m^2_{e\tau}/2K\ll 2\mu B$, the above expression for $P(\nu_{e}
\rightarrow \bar{\nu}_{\tau})$ then reduces to 
\begin{equation}\label{PSPVAC}
 P(\nu_{e}\rightarrow \bar{\nu}_{\tau})\simeq \sin^{2}(\mu B\Delta r) \,,
\end{equation}
which is independent of $K$.
Using the $B_{AGN}$ profile given by Eq.\ (\ref{magnet}) 
and for 
$\mu \sim 10^{-12}\mu_{B}$ ($\mu_{B}$ is Bohr magneton) \cite{R}, 
we obtain $P(\nu_{e}\rightarrow \bar{\nu}
_{\tau})>1/2$ for $\Delta m^{2}_{e\tau}\ll 10^{-2}$ eV$^{2}$. 
For this value of $\Delta m^{2}_{e\tau}$ ($\sim 10^{-3}$ eV$^{2}$),
the resonant spin-flavor precession due to matter induced gravity effects
 do not take place  
since the relevant level crossing condition is not satisfied.
In contrast to the vacuum flavor oscillations
mentioned earlier,
the value of $P$ in this case can be different
from 1/2, thus raising the possibility of distinguishing between the two
neutrino oscillation mechanisms. 
For  $\Delta m^2_{e\tau}/2K\sim  2\mu B$ or
$\Delta m^2_{e\tau}/2K\gg 2\mu B$,  
$P(\nu_{e}\rightarrow \bar{\nu}_{\tau})$ is $K$ dependent and is 
such that $P(\nu_{e}\rightarrow \bar{\nu}_{\tau})\leq 1/2$.

Taking into account now the effects of $\Delta V_{SF}$,
the relevant level crossing condition [$\Delta H_{SF}=0$, 
see Eq.\ (\ref{spin})] can be
satisfied for $\Delta m^{2}_{e\tau}\sim (10^{-8}-10^{-9})$ eV$^{2}$. This
relatively small value of $\Delta m^{2}_{e\tau}$ is also interesting 
in the context of supernova explosions\cite{SN} and the Sun\cite{GN}. 
Furthermore, this value of 
$\Delta m^{2}_{e\tau}$ is about 2 orders of magnitude 
larger than the one
required for flavor level crossing [see Eq.\ (\ref{flv})]. 
This
is due to the fact that the nucleon contribution in $\Delta V_{SF}$ does not
vanish (in contrast to that in $\Delta V_{F}$), as we emphasized earlier.

%
% fig 1
%
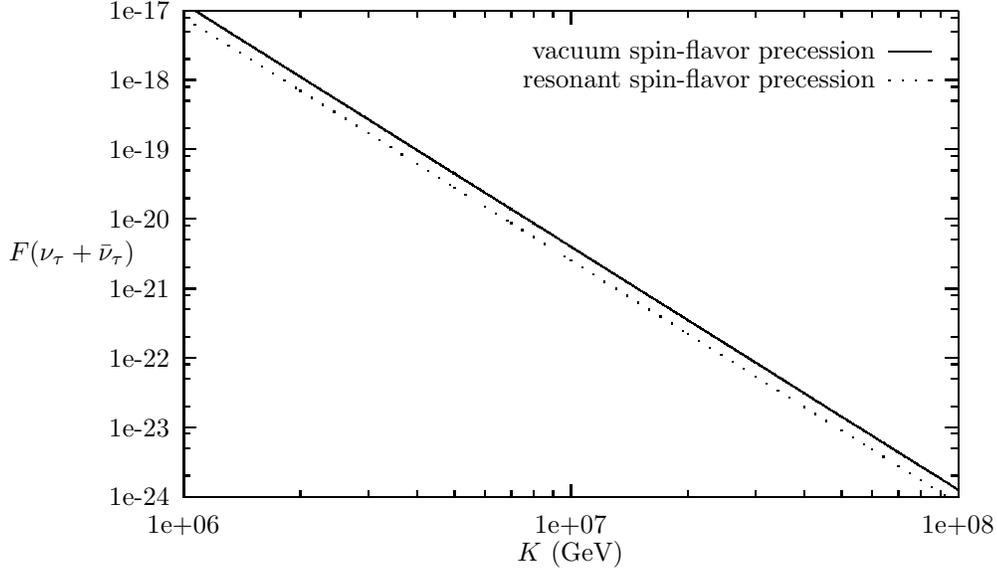
\begin{figure}
\begin{center}
% GNUPLOT: LaTeX picture
\setlength{\unitlength}{0.240900pt}
\ifx\plotpoint\undefined\newsavebox{\plotpoint}\fi
\sbox{\plotpoint}{\rule[-0.200pt]{0.400pt}{0.400pt}}%
\begin{picture}(1500,900)(0,0)
\font\gnuplot=cmr10 at 10pt
\gnuplot
\sbox{\plotpoint}{\rule[-0.200pt]{0.400pt}{0.400pt}}%
\put(220.0,113.0){\rule[-0.200pt]{4.818pt}{0.400pt}}
\put(198,113){\makebox(0,0)[r]{1e-24}}
\put(1416.0,113.0){\rule[-0.200pt]{4.818pt}{0.400pt}}
\put(220.0,146.0){\rule[-0.200pt]{2.409pt}{0.400pt}}
\put(1426.0,146.0){\rule[-0.200pt]{2.409pt}{0.400pt}}
\put(220.0,189.0){\rule[-0.200pt]{2.409pt}{0.400pt}}
\put(1426.0,189.0){\rule[-0.200pt]{2.409pt}{0.400pt}}
\put(220.0,212.0){\rule[-0.200pt]{2.409pt}{0.400pt}}
\put(1426.0,212.0){\rule[-0.200pt]{2.409pt}{0.400pt}}
\put(220.0,222.0){\rule[-0.200pt]{4.818pt}{0.400pt}}
\put(198,222){\makebox(0,0)[r]{1e-23}}
\put(1416.0,222.0){\rule[-0.200pt]{4.818pt}{0.400pt}}
\put(220.0,255.0){\rule[-0.200pt]{2.409pt}{0.400pt}}
\put(1426.0,255.0){\rule[-0.200pt]{2.409pt}{0.400pt}}
\put(220.0,298.0){\rule[-0.200pt]{2.409pt}{0.400pt}}
\put(1426.0,298.0){\rule[-0.200pt]{2.409pt}{0.400pt}}
\put(220.0,321.0){\rule[-0.200pt]{2.409pt}{0.400pt}}
\put(1426.0,321.0){\rule[-0.200pt]{2.409pt}{0.400pt}}
\put(220.0,331.0){\rule[-0.200pt]{4.818pt}{0.400pt}}
\put(198,331){\makebox(0,0)[r]{1e-22}}
\put(1416.0,331.0){\rule[-0.200pt]{4.818pt}{0.400pt}}
\put(220.0,364.0){\rule[-0.200pt]{2.409pt}{0.400pt}}
\put(1426.0,364.0){\rule[-0.200pt]{2.409pt}{0.400pt}}
\put(220.0,408.0){\rule[-0.200pt]{2.409pt}{0.400pt}}
\put(1426.0,408.0){\rule[-0.200pt]{2.409pt}{0.400pt}}
\put(220.0,430.0){\rule[-0.200pt]{2.409pt}{0.400pt}}
\put(1426.0,430.0){\rule[-0.200pt]{2.409pt}{0.400pt}}
\put(220.0,440.0){\rule[-0.200pt]{4.818pt}{0.400pt}}
\put(198,440){\makebox(0,0)[r]{1e-21}}
\put(1416.0,440.0){\rule[-0.200pt]{4.818pt}{0.400pt}}
\put(220.0,473.0){\rule[-0.200pt]{2.409pt}{0.400pt}}
\put(1426.0,473.0){\rule[-0.200pt]{2.409pt}{0.400pt}}
\put(220.0,517.0){\rule[-0.200pt]{2.409pt}{0.400pt}}
\put(1426.0,517.0){\rule[-0.200pt]{2.409pt}{0.400pt}}
\put(220.0,539.0){\rule[-0.200pt]{2.409pt}{0.400pt}}
\put(1426.0,539.0){\rule[-0.200pt]{2.409pt}{0.400pt}}
\put(220.0,550.0){\rule[-0.200pt]{4.818pt}{0.400pt}}
\put(198,550){\makebox(0,0)[r]{1e-20}}
\put(1416.0,550.0){\rule[-0.200pt]{4.818pt}{0.400pt}}
\put(220.0,582.0){\rule[-0.200pt]{2.409pt}{0.400pt}}
\put(1426.0,582.0){\rule[-0.200pt]{2.409pt}{0.400pt}}
\put(220.0,626.0){\rule[-0.200pt]{2.409pt}{0.400pt}}
\put(1426.0,626.0){\rule[-0.200pt]{2.409pt}{0.400pt}}
\put(220.0,648.0){\rule[-0.200pt]{2.409pt}{0.400pt}}
\put(1426.0,648.0){\rule[-0.200pt]{2.409pt}{0.400pt}}
\put(220.0,659.0){\rule[-0.200pt]{4.818pt}{0.400pt}}
\put(198,659){\makebox(0,0)[r]{1e-19}}
\put(1416.0,659.0){\rule[-0.200pt]{4.818pt}{0.400pt}}
\put(220.0,692.0){\rule[-0.200pt]{2.409pt}{0.400pt}}
\put(1426.0,692.0){\rule[-0.200pt]{2.409pt}{0.400pt}}
\put(220.0,735.0){\rule[-0.200pt]{2.409pt}{0.400pt}}
\put(1426.0,735.0){\rule[-0.200pt]{2.409pt}{0.400pt}}
\put(220.0,757.0){\rule[-0.200pt]{2.409pt}{0.400pt}}
\put(1426.0,757.0){\rule[-0.200pt]{2.409pt}{0.400pt}}
\put(220.0,768.0){\rule[-0.200pt]{4.818pt}{0.400pt}}
\put(198,768){\makebox(0,0)[r]{1e-18}}
\put(1416.0,768.0){\rule[-0.200pt]{4.818pt}{0.400pt}}
\put(220.0,801.0){\rule[-0.200pt]{2.409pt}{0.400pt}}
\put(1426.0,801.0){\rule[-0.200pt]{2.409pt}{0.400pt}}
\put(220.0,844.0){\rule[-0.200pt]{2.409pt}{0.400pt}}
\put(1426.0,844.0){\rule[-0.200pt]{2.409pt}{0.400pt}}
\put(220.0,866.0){\rule[-0.200pt]{2.409pt}{0.400pt}}
\put(1426.0,866.0){\rule[-0.200pt]{2.409pt}{0.400pt}}
\put(220.0,877.0){\rule[-0.200pt]{4.818pt}{0.400pt}}
\put(198,877){\makebox(0,0)[r]{1e-17}}
\put(1416.0,877.0){\rule[-0.200pt]{4.818pt}{0.400pt}}
\put(220.0,113.0){\rule[-0.200pt]{0.400pt}{4.818pt}}
\put(220,68){\makebox(0,0){1e+06}}
\put(220.0,857.0){\rule[-0.200pt]{0.400pt}{4.818pt}}
\put(403.0,113.0){\rule[-0.200pt]{0.400pt}{2.409pt}}
\put(403.0,867.0){\rule[-0.200pt]{0.400pt}{2.409pt}}
\put(510.0,113.0){\rule[-0.200pt]{0.400pt}{2.409pt}}
\put(510.0,867.0){\rule[-0.200pt]{0.400pt}{2.409pt}}
\put(586.0,113.0){\rule[-0.200pt]{0.400pt}{2.409pt}}
\put(586.0,867.0){\rule[-0.200pt]{0.400pt}{2.409pt}}
\put(645.0,113.0){\rule[-0.200pt]{0.400pt}{2.409pt}}
\put(645.0,867.0){\rule[-0.200pt]{0.400pt}{2.409pt}}
\put(693.0,113.0){\rule[-0.200pt]{0.400pt}{2.409pt}}
\put(693.0,867.0){\rule[-0.200pt]{0.400pt}{2.409pt}}
\put(734.0,113.0){\rule[-0.200pt]{0.400pt}{2.409pt}}
\put(734.0,867.0){\rule[-0.200pt]{0.400pt}{2.409pt}}
\put(769.0,113.0){\rule[-0.200pt]{0.400pt}{2.409pt}}
\put(769.0,867.0){\rule[-0.200pt]{0.400pt}{2.409pt}}
\put(800.0,113.0){\rule[-0.200pt]{0.400pt}{2.409pt}}
\put(800.0,867.0){\rule[-0.200pt]{0.400pt}{2.409pt}}
\put(828.0,113.0){\rule[-0.200pt]{0.400pt}{4.818pt}}
\put(828,68){\makebox(0,0){1e+07}}
\put(828.0,857.0){\rule[-0.200pt]{0.400pt}{4.818pt}}
\put(1011.0,113.0){\rule[-0.200pt]{0.400pt}{2.409pt}}
\put(1011.0,867.0){\rule[-0.200pt]{0.400pt}{2.409pt}}
\put(1118.0,113.0){\rule[-0.200pt]{0.400pt}{2.409pt}}
\put(1118.0,867.0){\rule[-0.200pt]{0.400pt}{2.409pt}}
\put(1194.0,113.0){\rule[-0.200pt]{0.400pt}{2.409pt}}
\put(1194.0,867.0){\rule[-0.200pt]{0.400pt}{2.409pt}}
\put(1253.0,113.0){\rule[-0.200pt]{0.400pt}{2.409pt}}
\put(1253.0,867.0){\rule[-0.200pt]{0.400pt}{2.409pt}}
\put(1301.0,113.0){\rule[-0.200pt]{0.400pt}{2.409pt}}
\put(1301.0,867.0){\rule[-0.200pt]{0.400pt}{2.409pt}}
\put(1342.0,113.0){\rule[-0.200pt]{0.400pt}{2.409pt}}
\put(1342.0,867.0){\rule[-0.200pt]{0.400pt}{2.409pt}}
\put(1377.0,113.0){\rule[-0.200pt]{0.400pt}{2.409pt}}
\put(1377.0,867.0){\rule[-0.200pt]{0.400pt}{2.409pt}}
\put(1408.0,113.0){\rule[-0.200pt]{0.400pt}{2.409pt}}
\put(1408.0,867.0){\rule[-0.200pt]{0.400pt}{2.409pt}}
\put(1436.0,113.0){\rule[-0.200pt]{0.400pt}{4.818pt}}
\put(1436,68){\makebox(0,0){1e+08}}
\put(1436.0,857.0){\rule[-0.200pt]{0.400pt}{4.818pt}}
\put(220.0,113.0){\rule[-0.200pt]{292.934pt}{0.400pt}}
\put(1436.0,113.0){\rule[-0.200pt]{0.400pt}{184.048pt}}
\put(220.0,877.0){\rule[-0.200pt]{292.934pt}{0.400pt}}
\put(45,495){\makebox(0,0){$F(\nu_{\tau}+\bar{\nu}_{\tau})$}}
\put(828,23){\makebox(0,0){$K$ (GeV)}}
\put(220.0,113.0){\rule[-0.200pt]{0.400pt}{184.048pt}}
\put(1306,812){\makebox(0,0)[r]{vacuum spin-flavor precession}}
\multiput(238.00,875.92)(0.794,-0.499){205}{\rule{0.735pt}{0.120pt}}
\multiput(238.00,876.17)(163.475,-104.000){2}{\rule{0.367pt}{0.400pt}}
\multiput(403.00,771.92)(0.799,-0.499){131}{\rule{0.739pt}{0.120pt}}
\multiput(403.00,772.17)(105.467,-67.000){2}{\rule{0.369pt}{0.400pt}}
\multiput(510.00,704.92)(0.793,-0.498){93}{\rule{0.733pt}{0.120pt}}
\multiput(510.00,705.17)(74.478,-48.000){2}{\rule{0.367pt}{0.400pt}}
\multiput(586.00,656.92)(0.799,-0.498){71}{\rule{0.738pt}{0.120pt}}
\multiput(586.00,657.17)(57.469,-37.000){2}{\rule{0.369pt}{0.400pt}}
\multiput(645.00,619.92)(0.802,-0.497){57}{\rule{0.740pt}{0.120pt}}
\multiput(645.00,620.17)(46.464,-30.000){2}{\rule{0.370pt}{0.400pt}}
\multiput(693.00,589.92)(0.791,-0.497){49}{\rule{0.731pt}{0.120pt}}
\multiput(693.00,590.17)(39.483,-26.000){2}{\rule{0.365pt}{0.400pt}}
\multiput(734.00,563.92)(0.799,-0.496){41}{\rule{0.736pt}{0.120pt}}
\multiput(734.00,564.17)(33.472,-22.000){2}{\rule{0.368pt}{0.400pt}}
\multiput(769.00,541.92)(0.820,-0.495){35}{\rule{0.753pt}{0.119pt}}
\multiput(769.00,542.17)(29.438,-19.000){2}{\rule{0.376pt}{0.400pt}}
\multiput(800.00,522.92)(0.781,-0.495){33}{\rule{0.722pt}{0.119pt}}
\multiput(800.00,523.17)(26.501,-18.000){2}{\rule{0.361pt}{0.400pt}}
\multiput(828.00,504.92)(0.796,-0.499){227}{\rule{0.737pt}{0.120pt}}
\multiput(828.00,505.17)(181.471,-115.000){2}{\rule{0.368pt}{0.400pt}}
\multiput(1011.00,389.92)(0.799,-0.499){131}{\rule{0.739pt}{0.120pt}}
\multiput(1011.00,390.17)(105.467,-67.000){2}{\rule{0.369pt}{0.400pt}}
\multiput(1118.00,322.92)(0.793,-0.498){93}{\rule{0.733pt}{0.120pt}}
\multiput(1118.00,323.17)(74.478,-48.000){2}{\rule{0.367pt}{0.400pt}}
\multiput(1194.00,274.92)(0.799,-0.498){71}{\rule{0.738pt}{0.120pt}}
\multiput(1194.00,275.17)(57.469,-37.000){2}{\rule{0.369pt}{0.400pt}}
\multiput(1253.00,237.92)(0.802,-0.497){57}{\rule{0.740pt}{0.120pt}}
\multiput(1253.00,238.17)(46.464,-30.000){2}{\rule{0.370pt}{0.400pt}}
\multiput(1301.00,207.92)(0.791,-0.497){49}{\rule{0.731pt}{0.120pt}}
\multiput(1301.00,208.17)(39.483,-26.000){2}{\rule{0.365pt}{0.400pt}}
\multiput(1342.00,181.92)(0.799,-0.496){41}{\rule{0.736pt}{0.120pt}}
\multiput(1342.00,182.17)(33.472,-22.000){2}{\rule{0.368pt}{0.400pt}}
\multiput(1377.00,159.92)(0.820,-0.495){35}{\rule{0.753pt}{0.119pt}}
\multiput(1377.00,160.17)(29.438,-19.000){2}{\rule{0.376pt}{0.400pt}}
\multiput(1408.00,140.92)(0.781,-0.495){33}{\rule{0.722pt}{0.119pt}}
\multiput(1408.00,141.17)(26.501,-18.000){2}{\rule{0.361pt}{0.400pt}}
\put(1328.0,812.0){\rule[-0.200pt]{15.899pt}{0.400pt}}
\put(1306,767){\makebox(0,0)[r]{resonant spin-flavor precession}}
\multiput(1328,767)(20.756,0.000){4}{\usebox{\plotpoint}}
\put(1394,767){\usebox{\plotpoint}}
\put(220,866){\usebox{\plotpoint}}
\multiput(220,866)(17.574,-11.044){11}{\usebox{\plotpoint}}
\multiput(403,751)(17.591,-11.015){6}{\usebox{\plotpoint}}
\multiput(510,684)(17.549,-11.083){4}{\usebox{\plotpoint}}
\multiput(586,636)(17.584,-11.027){4}{\usebox{\plotpoint}}
\multiput(645,599)(17.601,-11.000){2}{\usebox{\plotpoint}}
\multiput(693,569)(17.528,-11.115){3}{\usebox{\plotpoint}}
\multiput(734,543)(17.572,-11.045){2}{\usebox{\plotpoint}}
\multiput(769,521)(17.441,-11.252){2}{\usebox{\plotpoint}}
\put(817.43,490.42){\usebox{\plotpoint}}
\multiput(828,484)(17.574,-11.044){11}{\usebox{\plotpoint}}
\multiput(1011,369)(17.591,-11.015){6}{\usebox{\plotpoint}}
\multiput(1118,302)(17.549,-11.083){4}{\usebox{\plotpoint}}
\multiput(1194,254)(17.584,-11.027){3}{\usebox{\plotpoint}}
\multiput(1253,217)(17.601,-11.000){3}{\usebox{\plotpoint}}
\multiput(1301,187)(17.528,-11.115){2}{\usebox{\plotpoint}}
\multiput(1342,161)(17.572,-11.045){2}{\usebox{\plotpoint}}
\multiput(1377,139)(17.441,-11.252){2}{\usebox{\plotpoint}}
\put(1414.88,114.87){\usebox{\plotpoint}}
\put(1418,113){\usebox{\plotpoint}}
\end{picture}
\end{center}
\caption{$F(\nu_{\tau}+\bar{\nu}_{\tau})$ 
 [cm$^{-2}$s$^{-1}$sr$^{-1}$GeV$^{-1}$]
  Vs $K $(GeV). The solid curve is obtained for $P\sim 0.8$ 
 [using Eq.\ (\ref{PSPVAC})], whereas the dotted curve is obtained using 
 Eq.\ (\ref{PLZ}). In both cases, we use $\mu \sim 10^{-12} \mu_{B}$.
\label{fig:Fnutau}}
\end{figure}

The adiabaticity condition at the level crossing in this case is
\cite{marciano}
\begin{equation}\label{KSF}
 \kappa_{SF}\equiv \frac{2(2\mu B)^{2}}{|\Delta \dot{V}_{SF}|}\geq 1.
\end{equation}
Using the $B_{AGN}$ profile given by Eq.\ (\ref{magnet}), 
this condition can be satisfied
for $\mu \sim 10^{-12}\mu_{B}$, which is an  
order of magnitude smaller than the value that is needed in 
the context of the Sun\cite{GN}. The general expression for $P(\nu_{e}
\rightarrow \bar{\nu}_{\tau})$, including the possible
non-adiabatic effects ($\kappa_{SF}<1$) is given by \cite{Rin0}
\begin{equation}\label{PLZ}
 P(\nu_{e}\rightarrow \bar{\nu}_{\tau})=0.5-(0.5-P_{LZ})\cos2 \theta_{B},
\end{equation}
where $\tan2\theta_{B}=(-2\mu B)/\Delta H_{SF}$ is being evaluated at the 
neutrino production site and $P_{LZ}=\exp(-\frac{\pi}{2}\kappa_{SF})$. 
The expected $\nu_{\tau}$ flux spectrum due to  
(resonant) spin-flavor precession is calculated as
\begin{equation}\label{FTAU}
 F(\nu_{\tau}+\bar{\nu}_{\tau})=P(\nu_{e}\rightarrow \bar{\nu}_{\tau})
 F^{0}(\nu_{e}+\bar{\nu}_{e})+[1-P(\nu_{e}\rightarrow \bar{\nu}_{\tau})]
  F^{0}(\nu_{\tau}+\bar{\nu}_{\tau}).
\end{equation}
A similar expression for $F(\nu_{\tau}+\bar{\nu}_{\tau})$ due to 
flavor/spin-flavor oscillations in other channels can be straightforwardly
obtained with appropriate changes. For $F^{0}(\nu_{e}+\bar{\nu}_{e})$ 
we use the results from
Ref.\ \cite{SP} for $K \geq 10^{6}$ GeV as an example and, since 
$F^{0}(\nu_{\tau}+\bar{\nu}_{\tau})/F^{0}(\nu_{e,\mu}+
\bar{\nu}_{e,\mu})\ll 1$, the explicit form of $F^{0}(\nu_{\tau}+
\bar{\nu}_{\tau})$ is not important here. For illustrative purposes, 
in Fig. \ref{fig:Fnutau}, we display $F(\nu_{\tau}+
\bar{\nu}_{\tau})$ as a function of neutrino energy for
various neutrino spin-flavor transition mechanisms,
using Eq.\ (\ref{FTAU}). In the resonant spin-flavor precession
case, since the adiabaticity condition [Eq. (\ref{KSF})] is satisfied, 
we have $P_{LZ}\, \sim 0$. 
Also, since the corresponding value of $\cos 2\theta_{B}$ is quite 
small, $P(\nu_{e}\rightarrow \bar{\nu}_{\tau})\, \sim \, 1/2$ 
[see Eq. (\ref{PLZ})]. We have verified that
a similar behavior of
$P(\nu_{e}\rightarrow \bar{\nu}_{\tau})$ results
if the matter-induced 
gravity effects are omitted, but for 
a value of $\Delta m^{2}$ which, for the reason we
mentioned earlier, is approximately 2 
orders of magnitude smaller.

Another possibility, which we have not considered in detail,
is that the transitions occur between active and 
sterile neutrinos. 
As in the case just discussed, the nucleon contribution in the
Hamiltonian is important in this case also.
In the presence of a magnetic field,
the resonant spin-flavor 
transition may lead to energy dependent disappearance/appearance of the
active neutrino fluxes even if one uses the stringent astrophysical 
upper bounds on the relevant transition magnetic moment\cite{R}.

%
% section 5
%
\section{Conclusions}
\setcounter{equation}{0}
\label{sec:conclusions}

When neutrinos propagate in a medium
composed of electrons and nucleons,
their gravitational couplings are modified 
due to the weak interactions with the particles in the background.
While the tree level gravitational interactions of the neutrinos
have no effect on the oscillation phenomena, the matter-induced
gravitational couplings depend on the background particle number
densities and, in the presence of a gravitational potential,
they lead to additional contributions to the neutrino indices of refraction
which are not the same for the various neutrino flavors.

We have considered the
effects that such matter-induced couplings may have on the
phenomena associated with resonant neutrino oscillations,
under the combined presence of a gravitational and a magnetic field.
As an example of a setting where these effects are relevant,
we have studied their influence on the determination of the flux
of high energy
neutrinos from the core of Active Galactic Nuclei.
In that context 
we have pointed out that the resonant neutrino spin-flavor transitions may 
take place
for $\Delta m^{2} \sim (10^{-8}-10^{-9})$ eV$^{2}$
and $\mu \sim 10^{-12}\mu_{B}$, and
due to the matter-induced gravity effects the estimated high 
energy tau neutrino flux is somewhat higher than it would be if those 
effects are not included.
We have relied on the perturbative calculation of the
matter-induced gravitational couplings carried out in
Ref.\ \cite{np:gravnu}.
According to the current interpretation of the observed photon
flux from cores of AGN's, the neutrino production should take
place for values of $x$ around $10-100$ as we have assumed in the text. 
In this case, the use of those perturbative results is justified and
as a consequence the conditions for resonant transitions do
not depend on the choice of any metric parameters unlike in Ref. \cite{prw}. 

We have set aside some possible
incoherent effects which could be present for the high energy 
neutrinos that we are considering.
The considerations of the present work indicate
that  more detailed calculations taking them into account 
are worth pursuing, and that 
they have useful applications in the context of neutrino
emission in the vicinity of the core of AGN, with interesting
implications for km$^{2}$ surface area high energy neutrino telescopes.     

\paragraph*{Acknowledgments}

This work has been supported by a 
fellowship from the
Japan Society for the Promotion of 
Science (HA) and by the 
U.S. National Science Foundation Grant PHY-9900766 (JFN).

\appendix
\section*{Appendix}
\setcounter{equation}{0}
\section{Equation for the amplitudes}
\setcounter{equation}{0}

Here we derive Eq.\ (\ref{HamB}).  We begin by considering one
neutrino, with both a left- and a right-handed component, 
propagating through the medium with momentum 
$k^\mu = (\omega,\vec K)$,
in the presence of a gravitational potential and a magnetic field
$\vec B$. In a homogeneous medium,
the dispersion relation and wavefunction of the propagating modes
are determined from the linear part
of the effective field equation, 
which in momentum space takes the form 
\begin{equation}\label{fieldeq}
({k}\llap{/} - m - \Sigma_{eff} + \mu\vec\Sigma\cdot \vec B)\psi = 0 \,.
\end{equation}
$\Sigma_{eff}$ is the self-energy of the neutrino, which in the situation
that we are envisaging takes the form
\begin{equation}\label{sigmaeff}
\Sigma_{eff} = b_L{u}\llap{/} L + b_R{u}\llap{/} R\,,
\end{equation}
where
\begin{equation}\label{b}
b_L = b_{\mbox{mat}} + b_G \,,
\end{equation}
with $b_{\mbox{mat}}$ and $b_G$ being given by the formulas
quoted in Eqs.\ (\ref{bnuall}) and (\ref{bGfinal}). $u^\mu$ stands for the velocity
four-vector of the background medium. We adopt the frame in which
the background is at rest, so that $u^\mu = (1,\vec 0)$.

The formula for $b_R$ depends
on the nature of the right-handed component field.  
If it is the conjugate field of a standard left-handed neutrino,
then $b_R = -b_L$.  However, for a weak singlet field,
which we assume to be the case in what follows, $b_R = 0$.
In Eq.\ (\ref{sigmaeff}) we are neglecting terms of order $g^2/m^4_W$, as well
as the magnetic-field dependent term that arises from the effective
electromagnetic interactions of the neutrino in matter\cite{dnp}.
The latter contribution is not significant
for the values of the magnetic field that we are considering here.
In addition, we have dropped the purely gravitational term 
denoted by $b_g$ in Ref.\ \cite{np:gravnu}, which has a universal value
for all the neutrinos (including the sterile ones) and therefore
is not relevant for neutrino oscillations.

Writing 
\begin{equation}\label{psiweyl}
\psi = \left(
\begin{array}{c}
\xi \\
\eta
\end{array}
\right),
\end{equation}
in the Weyl representation of the gamma matrices, Eq.\ (\ref{fieldeq})
becomes
\begin{eqnarray}\label{weyleq}
(\omega - b_L + \vec\sigma\cdot\vec K)\eta - m\xi + 
\mu(\vec\sigma\cdot\vec B)\xi & = 0 \,,\nonumber\\
(\omega - \vec\sigma\cdot\vec K)\xi - m\eta + 
\mu(\vec\sigma\cdot\vec B)\eta & = 0 \,,
\end{eqnarray}
where $\vec\sigma$ are the Pauli matrices.  
Consider first the case in which $\vec B = 0$. In this case,
the equations have non-trivial solutions only if
$\xi$ and $\eta$ are proportional to the same spinor
$\phi_{\lambda}$ with definite helicity.
The positive energy solutions are then
$\omega_{-} \simeq K + b_L + \frac{m^2}{2K}$ and
$\omega_{+} \simeq K + \frac{m^2}{2K}$
for $\lambda = -1$ and $\lambda = +1$, respectively,
with the corresponding Dirac wavefunctions
\begin{eqnarray}\label{wavefuncB0}
\psi_{-} & = & \psi_L + y\left(
\begin{array}{cc}
\phi_{-} \\
0
\end{array}
\right), \nonumber\\
\psi_{+} & = & \psi_R + y^\prime\left(
\begin{array}{cc}
0 \\
\phi_{+}
\end{array}
\right),
\end{eqnarray}
where $y,y^\prime$ are of order $m/K$, and
\begin{equation}\label{chiralwavefunc}
\psi_L = \left(
\begin{array}{cc}
0 \\
\phi_{-}
\end{array}
\right) \,, \qquad
\psi_R = \left(
\begin{array}{cc}
\phi_{+} \\
0
\end{array}
\right) \,.
\end{equation}

When $\vec B \not = 0$, it is convenient to decompose $\vec B$ 
according to
\begin{equation}\label{splitB}
\vec B = B_{||}\hat K + \vec B_T \,.
\end{equation}
Then it is easy to see that, if $\vec B_T \not = 0$,
the two helicity spinors get mixed in Eq.\ (\ref{weyleq}),
and therefore we put
\begin{eqnarray}\label{ansatzB}
\eta & = & \alpha\phi_{-} + \epsilon\phi_{+} \,, \nonumber\\
\xi & = & \beta\phi_{+} + \epsilon^\prime\phi_{-} \,.
\end{eqnarray}
Substituting Eq.\ (\ref{ansatzB}) into Eq.\ (\ref{weyleq}) it follows that
$\epsilon$ and $\epsilon^\prime$ are of order
$m/K$ and $\mu B_{||}/K$,
%\begin{eqnarray}\label{epsilons}
%\epsilon & \simeq & \left(\frac{m - \mu B_{||}}
%{2K}\right)\beta \nonumber\\
%\epsilon^{\prime} & \simeq & \left(\frac{m + \mu B_{||}}
%{2K}\right)\alpha \,,
%\end{eqnarray}
%
while the equations that are obtained for $\alpha$ and $\beta$
can be recast in the form 
\begin{equation}\label{hameq}
H
\left(\begin{array}{cc}
\alpha\\
\beta
\end{array}\right) 
= \omega
\left(\begin{array}{cc}
\alpha\\
\beta
\end{array}\right) \,,
\end{equation}
where
\begin{equation}\label{HamBgen}
H = 
\left(\begin{array}{cc}
K + b_L + 
\frac{(m + \mu B_{||})^2}{2K} & -\mu B_T \\[12pt]
-\mu B_T & K + \frac{(m - \mu B_{||})^2}{2K}
\end{array} \,
\right).
\end{equation}
For a given
solution of Eq.\ (\ref{hameq}), the Dirac wavefunction of the
corresponding mode is obtained from Eqs.\ (\ref{psiweyl}) and (\ref{ansatzB}).
Neglecting terms proportional to $\epsilon$ and $\epsilon^\prime$,
it is then given by
\begin{equation}\label{diracsol}
\psi = \alpha\psi_L + \beta\psi_R \,,
\end{equation}
where $\psi_{L,R}$ are defined in Eq.\ (\ref{chiralwavefunc}).  
For an inhomogeneous medium, $H$ is taken as the Hamiltonian
for the amplitudes $\alpha,\beta$.

The generalization to two or more families is straightforward.  
The same procedure yields an equation for the amplitudes
that is identical to Eq.\ (\ref{hameq}), but where each element of 
$H$ is considered to be a matrix in the neutrino flavor space.
Similarly, the amplitudes $\alpha$ and $\beta$ are spinors 
in that flavor space, and they determine the spin and flavor
content of the corresponding propagating mode according
to Eq.\ (\ref{diracsol}).  These are the results quoted in Eqs.\ (\ref{Chi}) and (\ref{HamB}).

In the second case that we consider, specified by the interaction
Lagrangian terms given in Eqs.\ (\ref{etaumixing}) and (\ref{model2}),
the field equations, including the matter effects, are
\begin{eqnarray}\label{fieldeq2}
({k}\llap{/} - b^{(\nu_e)}_L{u}\llap{/})\nu_{eL} - m_{\nu_e} N_{eR}
+ \mu(\vec\Sigma\cdot\vec B)\nu^c_{\tau R} & = & 0, \nonumber\\
{k}\llap{/}N_{eR} - m_{\nu_e}\nu_{eL} & = & 0, \nonumber\\
({k}\llap{/} - b^{(\overline\nu_\tau)}_R{u}\llap{/})\nu^c_{\tau R} - 
m_{\nu_\tau} N^c_{\tau L}
+ \mu(\vec\Sigma\cdot\vec B)\nu_{eL} & = & 0, \nonumber\\
{k}\llap{/}N^c_{\tau L} - m_{\nu_\tau}\nu^c_{\tau R} & = & 0 \,.
\end{eqnarray}
where\cite{footnote2}
\begin{equation}\label{btaubar}
b^{(\overline\nu_\tau)}_R = 
- b^{(\nu_\tau)}_{\mbox{mat}} - b^{(\nu_\tau)}_G \,.
\end{equation}
Restricting ourselves for the moment to the situation
in which $B_{||} = 0$,
the positive energy solutions of these equations
are found by putting, in the Weyl representation,
\begin{eqnarray}\label{ansatz2}
\nu_{eL} & = & \alpha_{\nu_e}
\left(\begin{array}{c}0 \\ \phi_{-}\end{array}\right), \nonumber\\
N_{eR} & = & \epsilon \left(\begin{array}{c} \phi_{-} \\ 0 \end{array}\right),
\nonumber\\
\nu^c_{\tau R} & = & \beta_{\overline\nu_\tau}
\left(\begin{array}{c} \phi_{+} \\ 0 \end{array}\right),\nonumber\\
N^c_{\tau L} & = & \epsilon^\prime
\left(\begin{array}{c}0 \\ \phi_{+}\end{array}\right).
\end{eqnarray}
Substituting these forms into Eq.\ (\ref{fieldeq2}), this procedure yields
\begin{eqnarray}\label{small2}
\epsilon & \simeq & \frac{m_{\nu_e}}{2K}\alpha_{\nu_e}, \nonumber\\
\epsilon^\prime & \simeq & \frac{m_{\nu_\tau}}{2K}\beta_{\overline\nu_\tau} \,,
\end{eqnarray}
while the equations for $\alpha_{\nu_e}$ and $\beta_{\overline\nu_\tau}$ can
be written in the form
\begin{equation}\label{hameq2}
H\left(
\begin{array}{c}
\alpha_{\nu_{eL}} \\
\beta_{\overline\nu_{\tau R}}
\end{array}
\right) = \omega\left(
\begin{array}{c}
\alpha_{\nu_{eL}} \\
\beta_{\overline\nu_{\tau R}}
\end{array}
\right) \,,
\end{equation}
with
\begin{equation}\label{Hametau2}
H = 
\left(\begin{array}{cc}
K + b^{(\nu_e)}_{\mbox{mat}} + b^{(\nu_e)}_G + 
\frac{m_{\nu_e}^2}{2K}  & -\mu B_T \\[12pt]
-\mu B_T & K - b^{(\nu_\tau)}_{\mbox{mat}} - b^{(\nu_\tau)}_G + 
\frac{m_{\nu_\tau}^2}{2K}
\end{array}
\right) \,.
\end{equation}
When $B_{||} \not = 0$, the components of the wave function shown
in Eq.\ (\ref{ansatz2}) acquire an admixture of the opposite helicity spinor.
Allowing this and carrying through the same steps as before, 
the resulting equation for $\alpha_{\nu_e}$ and $\beta_{\overline\nu_\tau}$
can be written once again as in Eq.\ (\ref{hameq2}), but with $H$ given by the
formula quoted in Eq.\ (\ref{Hametau}).


\begin{thebibliography}{99}
\bibitem{footnote1} In this respect, our results differs in a crucial
way from that of Ref.\ \cite{prw}. There, the tree-level gravitational
couplings were found to yield different contributions to the
left and to the right-handed neutrino dispersion relations,
while the matter effects on the gravitational couplings
were completely neglected.  As explained here, we disagree with
that treatment.
\bibitem{footnote2} In general the coefficients $b^{(\nu_\alpha)}$ 
are functions of $\omega$ and $\vec K$. In those cases, the relation
is $b^{(\overline\nu_\alpha)}_R(\omega,\vec K) = 
-b^{(\nu_\alpha)}_L(-\omega,-\vec K)$.
\end{thebibliography}
\end{document}